\documentclass[12pt]{article}
\pdfoutput=1

\usepackage{algorithm}
\usepackage{algpseudocode}
\usepackage[sumlimits,]{amsmath}
\usepackage{amsfonts}
\usepackage{amsthm}
\usepackage{amssymb}
\usepackage{graphicx}
\usepackage{hyperref}
\usepackage{listings}

\def \dt { \,\mbox{d}t}

\def \du { \, \mbox{d}u}
\def \dx { \, \mbox{d}x}

\def \dz { \,\mbox{d}z}

\def \calK {\mathcal{K}}

\def \calN { {\mathcal N}}

\def \calV { \mathcal{V}}
\def \calW { \mathcal{W}}
\def \calX { \mathcal{X}}
\def \calY { \mathcal{Y}}
\def \calZ {\mathcal{Z}}
\def \diam {\mbox{diameter}}
\def \eps { {\varepsilon}}

\def \st {:\,}

\def \iid { {\mbox{i.i.d.}\,}}

\def \one { \mathbf{1}}  
\def \rmP {\mathrm{P}}

\def \E { {\mathbb E}}
\def \g { {\,|\,}}
\def \p {\partial}

\newcommand{\Exp}[1]{ \E\left\{ #1 \right\} }
\newcommand{\Exparg}[2]{ \E_{#1}\left\{ #2 \right\} }

\newcommand{\KL}[2]{ \mbox{KL}[#1 \,||\, #2]}   

\newcommand{\m}[1]{\texttt{#1}}

\def \Nat {{\mathbb N}}
\def \Rone { {\mathbb R}}

\def \Rd { {{\mathbb R}^d}}
\def \Rdxd { {{\mathbb R}^{d\times d}}}

\def \RM {{\mathbb R}^M}

\newcommand{\pushfwd}[1]{ {#1_{\#}}}

\newcommand{\TV}[2]{ \mbox{TV}[#1 \,||\, #2]}   

\def \catw { {\mbox{Categorical}(z_n; w_n)}}
\def \catwequal { {\mbox{Categorical}(z_n; w_n\equiv 1 / N)}}

\def \lognormal { {\mbox{Lognormal}}}
\def \sigmoid { {\mbox{Sigmoid}}}
\def \softmax { {\mbox{Softmax}}}
\def \union { {\mbox{Union}}}

\newtheorem{theorem}{Theorem}[section]
\newtheorem*{theorem*}{Theorem}
\newtheorem{proposition}{Proposition}[section]

\theoremstyle{definition}
\newtheorem*{def*}{Definition}
\newtheorem{definition}{Definition}[section]

\theoremstyle{remark}
\newtheorem*{remark*}{Remark}

\newtheorem*{example*}{Example}

\newtheorem*{claim*}{Claim}

\title{Quadrature Compound:  An approximating family of distributions}
\author{J. Dillon \qquad I. Langmore \thanks{Google Inc.  Both authors contributed equally.}}

\begin{document}
\maketitle
\abstract{Compound distributions allow construction of a rich set of distributions.  Typically they involve an intractable integral.  Here we use a quadrature approximation to that integral to define the quadrature compound family.  Special care is taken that this approximation is suitable for computation of gradients with respect to distribution parameters.  This technique is applied to discrete (Poisson LogNormal) and continuous distributions.  In the continuous case, quadrature compound family naturally makes use of parameterized transformations of unparameterized distributions (a.k.a ``reparameterization''), allowing for gradients of expectations to be estimated as the gradient of a sample mean.  This is demonstrated in a novel distribution, the diffeomixture, which is is a reparameterizable approximation to a mixture distribution.}


\section{Introduction}
\label{section:introduction}

Given a marginal density $p(z)$, and conditional $p(x\g z)$, the \emph{compound distribution} $p(x)$ is defined by
\begin{align}
  \label{align:compound}
  p(x) :&= \int p(x\g z) p(z)\dz.
\end{align}
In the case that $X$ is continuous, $p(x\g z)$ is a conditional density, and in case $X$ is discrete, $p(x\g z) = \rmP[X=x\g Z=z]$ is a conditional probability.  This leads to $p(x)$ being a probability or probability density, which we collectively refer to as a probability function.

As an example, consider using the LogNormal $p(z)$ to parameterize the rate of the Poisson $p(x\g z)$:
\begin{align}
  \label{align:poisson-lognormal}
  \begin{split}
    p(x\g z) :&= \frac{z^x \exp\left\{ -z \right\}}{x!},
    \quad p(z) := \frac{1}{z\sigma\sqrt{2\pi}} \exp\left\{ -\frac{(\log z - \mu)^2}{2\sigma^2} \right\} \\
    p(x) &= \rmP[X=x]
    =\int
    \frac{z^x \exp\left\{ -z \right\}}{x!}
    \frac{1}{z\sigma\sqrt{2\pi}} \exp\left\{ -\frac{(\log z - \mu)^2}{2\sigma^2} \right\} \dz
  \end{split}
\end{align}
This is an attractive enhancement to the Poisson, since the two parameters $(\mu, \sigma)$ allow independent control over the mean and variance.
Unfortunately, the integral defining $p(x)$ is not available in closed form.

A closed for expression for compound distributions can be achieved in the special case of conjugate pairs~\cite{jordan-conjugate}.  For example, a parameterizing a Poisson's rate with a gamma (rather than the lognormal as above) yields the \emph{negative binomial distribution}.  For a continuous example consider parameterizing the variance of a Normal with an inverse gamma, leading to a \emph{non-standardized Student's T}.
Although tractable, conjugate pairs may be incompatible with the practicioners beliefs about the sytem at hand.  See~\cite{Blei06correlatedtopic} for an example where a softmax-normal is used to parameterize a (non-conjugate) multinomial, which allowed modeling of correlations and improved upon the conjugate pairing of Dirichlet + multinomial.  Conjugate pairs (such as the non-standardized Student's T) may also not possess additional desirable properties, such as reparameterizability, which complicates computation of gradients of expectations (section~\ref{section:transformed-distributions}).

Much work has been done on finding quadrature rules suitable to approximate intractable integrals such as \eqref{align:compound}.  The technique of Gaussian quadrature allows computation of expectations against many common probability functions~\cite{scipy-special,stoer-numerical-analysis}, and can be extended to arbitrary measures ~\cite{fernandes-gauss-arbitrary-positive}.  Gaussian (and other) quadrature techniques could be used to approximate the integral defining $p(x)$.  Sampling from $p(x)$ could then be performed exactly by drawing $z\sim p(z)$, then setting $x\sim p(x\g z)$.  However, this leaves one in a situation where the distribution function $p(x)$ does not exactly coincide with the samples $X$.

\begin{figure}
  \centering
  \includegraphics[width=0.46\textwidth]{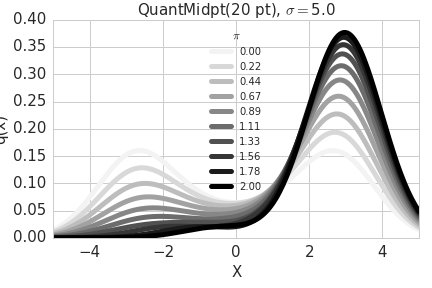}
  \includegraphics[width=0.46\textwidth]{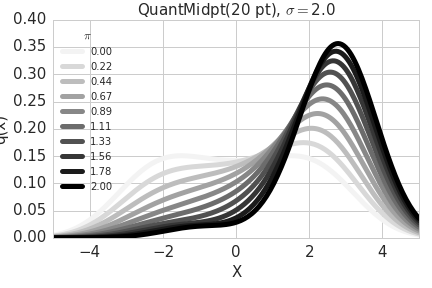}
  \caption{Two diffeomixture densities, both using the SigmoidNormal mixture density $p(z) = p(z;\pi,\sigma)$, and Normal components.  The component choice parameter $\pi$ is swept from 0 to 2.  Left:  Scale parameter $\sigma=5$, giving a pdf similar to a standard mixture, with the right component getting most of the weight as $\pi\to2$.  Right:  $\sigma=2$, which allows for a ``blend'' of the two components for smaller $\pi$.}
  \label{fig:vdm-smooth}
\end{figure}

\begin{figure}
  \centering
  \includegraphics[width=0.31\textwidth]{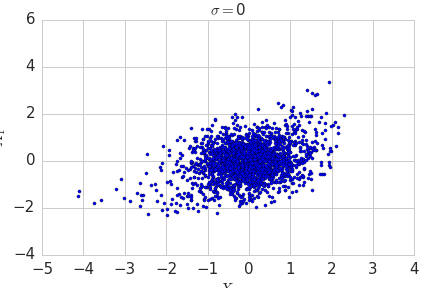}
  \includegraphics[width=0.31\textwidth]{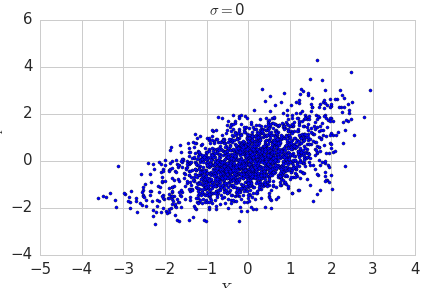}
  \includegraphics[width=0.31\textwidth]{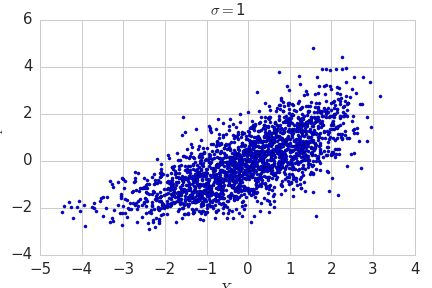} \\
  \includegraphics[width=0.31\textwidth]{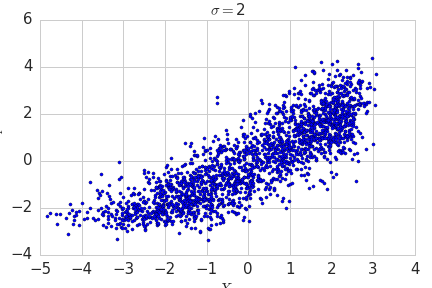}
  \includegraphics[width=0.31\textwidth]{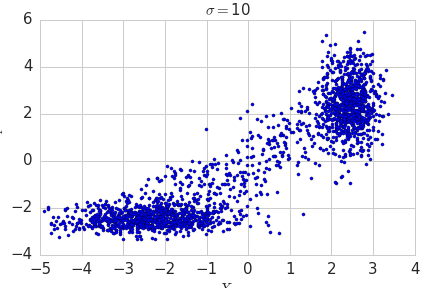}
  \includegraphics[width=0.31\textwidth]{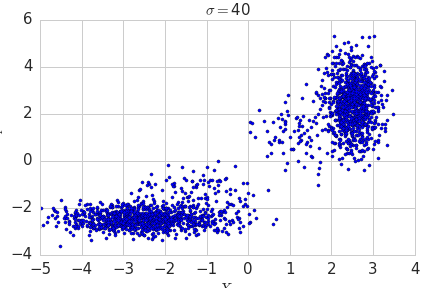}
  \caption{Sequence of samples from bi-mixtures (of two 2D Normals) showing a smooth transition as the mixture scale $\sigma$ varies from 0.1 (upper left) to 40 (lower right).}
  \label{fig:vdm-2d-smooth}
\end{figure}

The contribution of this work is to replace \eqref{align:compound} with a quadrature approximation $\sum_n p(x\g z_n) w_n$.  This summation \emph{defines} the quadrature compound (QC) as a mixture with special relationship between $z_n$ and $w_n$.  This allows the definition of samples and distribution function to coincide exactly, regardless of how few quadrature points are used.  If the quadrature approximation is done with care, desirable differentiability properties emerge.  For example, in the case of the Poisson LogNormal \eqref{align:poisson-lognormal}, the scheme allows gradients of $p(x)$ with respect to the parameters $(\mu, \sigma)$ to be taken (figure~\ref{fig:poisson-lognormal}).  A continuous example is also provided, dubbed the ``diffeomixture.''  This diffeomixture is a reparameterizable distribution that can approximate a standard mixture, or a ``smearing'' of the components (see figures~\ref{fig:vdm-smooth},\ref{fig:vdm-2d-smooth}).
The quadrature compound can thus be seen as extending the menu of computationally tractable compound distributions far beyond conjugate pairs.

This work develops the theory of QC distributions in arbitrary dimensions.  Code snippets demonstrate the instantiation of two QC distributions currently available in the TensorFlow library~\cite{tensorflow} library.  Both of these distributions are defined such that the integral \eqref{align:compound} is over a one dimensional variable $z$.  Extending to $M$ dimensions would require \emph{a-priori} exponentially more (in $M$) quadrature points.  This brings up the need for an in-depth look at higher dimensional quadrature, which should be considered in subsequent work.

The remainder of the paper is organized as follows.  Section~\ref{section:qc-trick} describes the approximation of \eqref{align:compound} with a quadrature scheme, detailing the schemes we found most useful.  Sections~\ref{section:poisson-lognormal-qc} and~\ref{section:reparameterizable-qc-and-diffeomixture} give concrete examples of discrete and continuous quadrature compounds, along with TensorFlow code snippets.  Section~\ref{section:transformed-distributions} describes how to define new distributions as diffeomorphic transformations of other distributions.  This is important for our discussion of reparameterizable distributions, and will be a (short, rigorous) restatement of one key approach to stochastic gradient estimation (see also e.g. ~\cite{fu-gradient-estimation,Schulman-2015-stochastic-computation-graphs,kingma-welling}).  Section~\ref{section:approximation} comprises the second half of this paper, giving detailed comparison of five quadrature schemes, as well as convergence proofs.  This may be ignored by those only interested in higher level details.

\section{The Quadrature Compound ``Trick''}
\label{section:qc-trick}
An obvious approximation for \eqref{align:compound} is the Monte Carlo sum
\begin{align*}
 p(x) \approx \frac{1}{N}\sum_{n=1}^N p(x\g z_n), \quad z_n\sim p(z).
\end{align*}
While this sum is an unbiased estimate of $p(x)$, for purposes of estimation we are often more interested in $\log p(x)$, and the logarithm of the above sum is \emph{not} an unbiased estiamte of $\log p(x)$.  One corrective approach would be to use a cautiously large $N$ and/or an importance sampling scheme.  Instead of Monte Carlo, our approach is to use a quadrature approximation to the integral over $z$.

\begin{definition}[Quadrature Scheme]
  \label{definition:quadrature-scheme}
  Given probability function $p(z)$ supported on metric space $\calZ$, we call the sequence of points/weights $\calW_N := \left\{ (z_{1,N}, w_{1,N}),\ldots,(z_{N,N}, w_{N,N}) \right\}$ defined for $N = 1,2,\ldots$ a \emph{quadrature scheme} for $p(z)$ if for every $N$, $w_{n, N}\geq0$, $\sum_{n=1}^N w_{n,N} = 1$, and given uniformly continuous and bounded $\varphi:\calZ\to\Rone$, we have
  \begin{align*}
    \lim_{N\to\infty} \sum_{n=1}^N w_{n,N} \varphi(z_{n,N}) = \int \varphi(z) p(z)\dz.
  \end{align*}
\end{definition}

Many quadrature schemes are possible, and here we present three that define $z_n$ as midpoints of quantiles (or multi-dimensional generalizations thereof) since this is simple and leads to constant weights, which are well-suited for creating reparameterizable samples (section~\ref{section:transformed-distributions}).  These schemes are defined for non-vanishing $p(z)$ with varying degree of regularity.  The nonvanishing requirement is there to prevent quantiles from stretching across disconnected portions of the support, and could be replaced by ``vanishing at a finite number of points'', or even a smoothness requirement on $p(z)$.  See section~\ref{section:approximation} for convergence proofs and consideration of other schemes.

\begin{definition}[Quantile midpoint scheme on bounded intervals]
  \label{definition:midpoint-quad-bounded}
  Suppose $p(z)$ is supported and non-vanishing on $\calZ\subset\Rone$, a bounded interval.
  Let $\nu_0,\ldots,\nu_N$ be the $n/N$ quantile of $p(z)$.  That is, $\rmP[Z\leq \nu_n] = n/N$.  Then $w_{n,N} \equiv 1/N$, and $z_n := (\nu_{n-1} + \nu_n)/2$, defines a quadrature scheme for $p(z)$.
\end{definition}

\begin{definition}[Quantile midpoint scheme on $[0, \infty)$.]
  \label{definition:midpoint-quad-half-bounded}
  Suppose $p(z)$ is supported and non-vanishing on $[0, \infty)$.
  Let $\nu_0,\ldots,\nu_{N-1}$ be the $n/N$ quantile of $p(z)$, and $\nu_N := \nu_{N-1} + (\nu_{N-1} - \nu_{N-2})$.  Then $w_{n,N} \equiv 1/N$, and $z_n := (\nu_{n-1} + \nu_n)/2$, defines a quadrature scheme for $p(z)$.  See section~\ref{subsection:quantile-convergence}.
\end{definition}

A multidimensional generalization of schemes~\ref{definition:midpoint-quad-bounded}~\ref{definition:midpoint-quad-half-bounded} is
\begin{definition}[Cubature constant-probability scheme]
  \label{definition:cubature-constant-probability-point-scheme}
  Suppose we have a non-vanishing probability density $p(z)$ defined on a metric space $\calZ$, and that for $N = 1,2,\ldots$, there exists a partition $\calZ = \delta_{1,N}\cup\cdots\cup\delta_{n,N}$ into regions such that $\rmP[\delta_{N,n}] \equiv 1 / N$.  Suppose further that for every $D>0$, the subset of $\delta_{N,n}$ with large diameter,
  \begin{align*}
    B_{D,N} :&= \union{\left\{ \delta_{n,N}\st \diam(\delta_{n,N}) > D \right\}},
  \end{align*}
  tends to zero in measure; that is, $\rmP[Z\in B_{D,N}]\to0$ as $N\to\infty$ for every $D>0$.
  Then we may take $w_{n,N} \equiv 1/N$, and $z_{n,N}$, to be any point interior to $\delta_{n,N}$, and we have a quadrature scheme for $p(z)$.  See section~\ref{subsection:cubature-convergence}.
\end{definition}

\begin{definition}[Quadrature Compound (QC)]
  We define the \emph{quadrature compound} distributional family as distributions of the form
\begin{align}
  \label{align:quad-compound}
  q_N(x) :&= \sum_{n=1}^N w_n p(x\g z_n),
\end{align}
  where $\calW_N$ is a quadrature scheme for some marginal density $p(z)$.
\end{definition}

Note that the conditions on $w_n$ ensure that $q(x)$ is non-negative and integrates/sums to one for every $N$, and  thus defines a probability function \emph{regardless of how well the quadrature approximation works}.
In fact, a look at \eqref{align:quad-compound} reveals the quadrature compound is a mixture distribution, with parameterized components $z_n$ and weights $w_n$.  This means that sampling from a QC can be done in two steps:  First, $z_n$ is drawn from $\catw$, a categorical that chooses $z_n\in\left\{ z_1,\ldots,z_N \right\}$ with probability $w_n$.  Second, $x$ is drawn from the conditional $p(x\g z_n)$.  It is also natural to compare a QC to a mixture with $N$ components.  The mixture will be more flexible, and the QC will ensure $q_N(x)\approx p(x)$.  So the user should choose a QC when she has an \emph{a-priori} belief about the applicability of $p(x)$.  For example, the mixture depends on $N$ parameters $z_n$ (and possibly $N$ more if $w_n$ are variable), while the $z_n$ could depend on far fewer (our examples have one or two).  The user may prefer this lower dimensional parameterization that forces a specific parameterized shape.

\section{Discrete Quadrature Compounds}
\label{section:poisson-lognormal-qc}
Using quantile midpoint quadrature scheme \ref{definition:midpoint-quad-half-bounded} for $p(z;\mu,\sigma) = \lognormal(\mu, \sigma)$, and $p(z\g x)$ the Poisson mass function, the Poisson-LogNormal \eqref{align:poisson-lognormal} can be approximated as the QC density
\begin{align*}
  q(x)
  &= \sum_{n=1}^N p(x \g z_n)w_n
  = \sum_{n=1}^N \frac{z_n^x \exp\left\{ -z_n \right\}}{x!} \frac{1}{N} ,\quad x = 0,1,2,\ldots.
\end{align*}
Since $z_n$ are functions of quantiles of $\lognormal(\mu,\sigma)$, they are differentiable with respect to $(\mu,\sigma)$.  Therefore, $\nabla_{\mu,\sigma} q(x)$ is analytically computable, which is advantageous for gradient based estimation of $(\mu,\sigma)$.

Samples $X$ are generated according to
\begin{enumerate}
  \item Draw $Z\sim\catwequal$
  \item Draw $X\sim\mathrm{Poisson}(Z)$
\end{enumerate}

\begin{figure}
  \centering
  \includegraphics[width=0.46\textwidth]{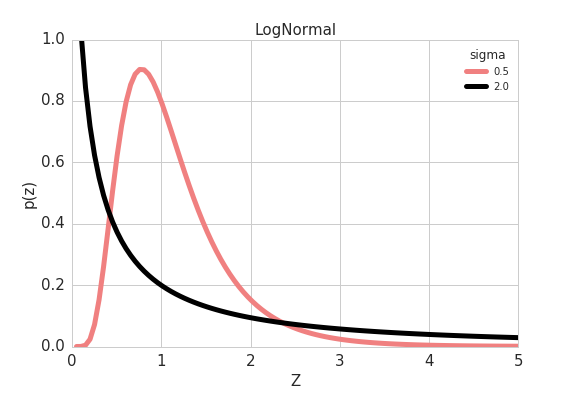}
  \includegraphics[width=0.46\textwidth]{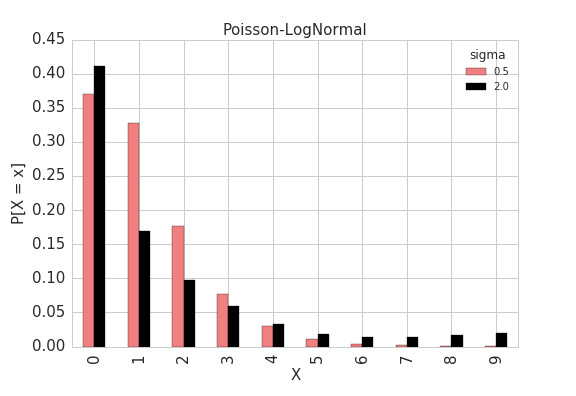}
  \caption{Left:  LogNormal density, with $\mu=0$, $\sigma\in\left\{ 0.5, 2.0 \right\}$.  Right:  Poisson-LogNormal quadrature compound with same $(\mu, \sigma)$.}
  \label{fig:poisson-lognormal}
\end{figure}

Figure~\ref{fig:poisson-lognormal} shows $\rmP[X = x]$ for two different values of $\sigma$.  Since $Z$ parameterizes the mean of the conditional Poisson density, places where one LogNormal curve is larger/smaller than the other roughly correspond to places where the same Poisson-LogNormal is larger/smaller.

The Poisson-LogNormal that generated figure~\ref{fig:poisson-lognormal} has been added to TensorFlow (see listing~\ref{listing:poisson-lognormal}).
\begin{lstlisting}[language=Python, caption=Poisson-LogNormal, label={listing:poisson-lognormal}, float=h]
  import tensorflow as tf
  ds = tf.contrib.distributions

  plognormal = ds.PoissonLogNormalQuadratureCompound(
      loc=0., scale=[0.5, 2.0])
\end{lstlisting}

Other discrete distributions can be handled in a manner similar to the Poisson-LogNormal QC.

\section{Reparameterization of Distributions}
\label{section:transformed-distributions}

The distributions defined in this paper are based on parameterized transformations of unparameterized distributions, a.k.a \emph{reparmaeterized}.  Here we review this process and its relationship to stochastic gradient estimation and optimization, stating a variation of well-known conditions under the naive stochastic gradient estimate is justified (theorem~\ref{theorem:reparameterizable-conditions}).  See \cite{fu-gradient-estimation} for a thorough discussion, or \cite{kingma-welling, Schulman-2015-stochastic-computation-graphs} for a discussion in the context of machine learning (where the term \emph{reparameterization trick} arose).

First, as an important example of reparameterization we consider diffeomorphisms.  Given manifolds $\calX, \calY$, we define a \emph{diffeomorphsim} $F:\calX\to\calY$ to be a bijective (one-to-one and onto) map such that the matrix of partial derivaties $DF_{ij} := \p F_i/\p x_j$ is continuous.  As a consequence of the inverse function theorem, $DF(y)^{-1}$ exists for all $y\in\calY$ and is continuous as well.
If $X$ is a continuous random variable supported in $\calX$, one may use the diffeomorphism to define $Y = F(X)$.  The probability density of $Y$, $p_Y$, can then be written in terms of a \emph{pushforward} of the density of $X$, $p_X$.
\begin{align}
  p_Y(y) &= (\pushfwd{F}p_X)(y) = p_X(F^{-1}(y))|DF^{-1}(y)|,
  \label{align:pushforward-density}
\end{align}
where above $|\cdot|$ is the absolute value of the determinant.  Note the notational correspondence $Y = F(X)\leftrightarrow p_Y = \pushfwd{F}p_X$.
Due to the explicit formula \eqref{align:pushforward-density}, a diffeomorphism is an especially convenient way to create a reparameterized distribution $p_Y$.

Reparameterization can be placed in the context of stochastic optimization by assuming $X$ is a random variable, $Y = F(X) = F(X;\lambda)$ is a parameterized transformation of $X$, and we wish to choose $\lambda$ to minimize $\Exp{\varphi(Y)}$ for a loss function $\varphi$.  $F$ does not have to be a diffeomorphism.
A common attack is to use a Robbins-Monro type stochastic optimization scheme, meaning we update $\lambda_j$ to $\lambda_{j+1} := \lambda_j - \tau_j G(\lambda_j)$, for some sequence $\tau_j\searrow0$ and $G(\lambda)$ an unbiased and sufficiently accurate estimate of $\p_\lambda\Exp{\varphi(Y)}$ \cite{Robbins&Monro:1951}.

The question we delve deeper into is how to approximate $\p_\lambda\Exp{\varphi(Y)}$.  The naive guess is to set
\begin{align*}
  S_K :&= \frac{1}{K} \sum_{k=1}^K Y_k = \frac{1}{K}\sum_{k=1}^K F(X_k, \lambda),\quad X_k\sim p_X,
\end{align*}
and since (under appropriate conditions) $S_K\to \Exp{\varphi(Y)}$, we also hope $\p_\lambda S_K\to \p_\lambda \Exp{\varphi(Y)}$.
If this is the case, we are justified using $S_K$ as an estimate of the loss $\Exp{\varphi(Y)}$, then letting auto-differentiation software blindly compute the gradient $\p_\lambda S_K$, so that it may be used in a gradient descent scheme.

The difficulty is that we cannot always exchange differentiation and integration.
A common failure mode is when samples $Y$ are generated according to some non-differentiable algorithm.  This prevents the most common Gamma sampling scheme from being reparameterizable with respect to its shape and scale parameters\cite{naesseth-reparameterizable-rejection}.  Consider also $X$ Normal, then if $\varphi(F(x)) = H(x - \lambda)$ is the step function centered at $\lambda$, one can check that $\p_\lambda S_K\equiv0$ almost surely, which is not equal to $\p_\lambda\Exp{H(X - \lambda)}$, rendering our finite sample estimator useless in a gradient based scheme.  A similar trouble plagues any jump discontinuity in $\varphi\circ F$.  By comparison, if $\varphi\circ F$ is smooth and bounded, theorem~\ref{theorem:reparameterizable-conditions} will apply.  Some non-smooth unbounded cases work as well, for example $\varphi\circ F(x) = |x - \lambda|$.

Let us state the main regularity condition, which is a Lipshitz condition, where the Lipshitz ``constant'' depends on $X$ and has finite expectation.
\begin{definition}
  \label{definition:local-lipshitz}
  We say that function $f:\calX\times\Rone\to\Rone$ is $L(x)-$Lipshitz if 
  \begin{align*}
    \left|\frac{f(x, \lambda + h) - f(x, \lambda) }{h}\right| & \leq L(x),
  \end{align*}
  for non-negative function $L$, with $\Exp{L(X)} < \infty$.
\end{definition}
An $L(x)-$Lipshitz function will be absolutely continuous in $\lambda$, which means the partial derivative $\p_\lambda(\varphi\circ F)$ exists as an $L^1$ function, bounded in absolute value by $L(x)$ \cite{folland}.
\begin{theorem}[Convergence of reparameterized gradient estimator]
  \label{theorem:reparameterizable-conditions}
  Let $X\in\calX$ be a continuous random variable with density $p_X$ independent of $\lambda\in\Rone$.  Set $Y = F(X) = F(X;\lambda)$.  Suppose we have function $\varphi$ such that $\Exp{|\varphi(F(X))|}<\infty$, and $\varphi\circ F$ is $L(x)-$Lipshitz.
  Then,
  \begin{align*}
    \Exp{S_K} &= \Exp{\varphi(Y)},\quad\mbox{and}\quad\Exp{\p_\lambda S_K} = \p_\lambda\Exp{\varphi(Y)},
  \end{align*}
  and as $K\to\infty$,
  \begin{align*}
    S_K\to \Exp{\varphi(Y)},\quad\mbox{and}\quad \p_\lambda S_K\to \p_\lambda \Exp{\varphi(Y)},
  \end{align*}
  almost surely.
\end{theorem}
\begin{proof}
  $\varphi(F(X))$ is clearly an unbiased estimate of $\Exp{\varphi(F(X))}$.  This and the assumption $\Exp{|\varphi(F(X))|}< \infty$ means $S_K\to\Exp{\varphi(Y)}$ by the strong law of large numbers.  Next,
  \begin{align*}
    \p_\lambda \Exparg{p_Y}{\varphi(Y)}
    &= \p_\lambda \Exparg{p_X}{\varphi(F(X))} \\
    &= \lim_{h\to0} \int\frac{\varphi(F(x, \lambda + h)) - \varphi(F(x, \lambda))}{h} p_X(x)\dx \\
    &= \Exparg{p_X}{\p_\lambda\varphi(F(X))},
  \end{align*}
  where the exchange of limit $h\to0$ and integration is justified using dominated convergence with dominating function $L(x)$.  This implies $\p_\lambda\varphi(F(X))$ is an unbiased estimate of $\p_\lambda\Exp{\varphi(Y)}$.  Since $\Exp{|\p_\lambda\varphi(F(X))}| \leq \Exp{L(X)} < \infty$, the strong law of large numbers implies $\p_\lambda S_K\to \p_\lambda\Exp{\varphi(Y)}$ as desired.
\end{proof}

Transformations of this type have been packaged into the \m{distributions} library of \m{TensorFlow} \cite{tensorflow}.  See listing~\ref{listing:sigmoid-normal} for a code snippet demonstrating $Z = F(U; \pi,\sigma) = 1 / (1 + \exp\left\{ \sigma\pi + \sigma U\right\})$.
\begin{lstlisting}[language=Python, caption=Sigmoid-Normal, label={listing:sigmoid-normal}, float=h]
  import tensorflow as tf
  ds = tf.contrib.distributions

  # sigma is positive and trainable
  sigma = tf.nn.softplus(tf.Variable(0.0))
  pi = 0.5

  # Make F(U) = 1 / (1 + Exp{-(sigma * pi + sigma * U)})
  sigmoid = ds.bijectors.Sigmoid()
  shift_and_scale = ds.bijectors.Affine(
      shift=sigma * pi, scale_identity_multiplier=sigma,
      event_ndims=0)
  F = ds.bijectors.Chain([sigmoid, shift_and_scale])

  Z = ds.TransformedDistribution(distribution=U, bijector=F)
\end{lstlisting}

\section{Reparameterizable Quadrature Compounds and the Diffeomixture}
\label{section:reparameterizable-qc-and-diffeomixture}
Here we discuss continuous QC distributions that are reparameterizable with respect to a parameter $\lambda$.  This means they can take advantage of theorem~\ref{theorem:reparameterizable-conditions} to compute gradients of expectations.  This puts two major limitations on the QC.  First, samples $X\sim p(x\g z)$ must be sufficiently smooth with respect to $\lambda$, and second, the weights $w_n$ must be independent of $\lambda$.  The first requirement means discrete distributions such as the Poisson-LogNormal will not be reparameterizable (see~\cite{bbvi} for alternative gradient schemes).  The second means many adaptive quadrature schemes cannot be used.

Our application is the diffeomixture, which approximates a mixture with a reparameterizable QC.  To see the connection between mixtures and compound distributions, we will write a (standard) mixture in the form of a compound distribution.  Let $p(z) = \sum_m \pi_m\delta(z - z_m)$, where $\delta(z - z_m)$ is the dirac mass centered at $z_m$.  Then
\begin{align*}
  p(x) = \int p(x\g z) \sum_m\pi_m\delta(z - z_m) = \sum_m \pi_m p(x\g z_m)
\end{align*}
is a mixture with mixture weights $\pi_m$.  A diffeomixture relaxes this in two ways.  First, we approximate $\sum_m\pi_m\delta(z - z_m)$ with a smooth function.  Second, we use the quadrature trick to approximate the integral.

\subsection{Softmax and sigmoid mixture weights}
\label{subsection:sigmoid-softmax-mixture-weights}
Here we define our smooth substitute for $\sum_m\pi_m\delta(z - z_m)$, which is a generalization of the gumbel-softmax/concrete distribution~\cite{gumbel-softmax,concrete} with a slightly different parameterization.
This distribution relies on a softmax transformation, and is dubbed the \emph{softmax mixture weight}.  
The softmax allows us to define densities $p(z)$ over the M-simplex
\begin{align*}
\Delta^M := \left\{ (t_0,\ldots,t_M): t_m\geq 0, \sum_m t_m = 1 \right\},
\end{align*}
which sum to one and thus are a continuous relaxation of mixture weights.  In this case, the integral $p(x) = \int p(x\g z)p(z)\dz$ is over the M-simplex, and the measure $\dz$ is the measure on the simplex, \emph{not} Lebesgue measure on $\RM$.  The apparent technical difficulty of integrating against a non-Lebesgue measure is avoided, since our QC performs this integral with a summation, and the associated convergence proof only requires that $\calZ$ is a metric space.  In the case of two mixture weights ($M=2$), the technicality can be further avoided by identifying $\Delta^M$ with the interval $[0, 1]$, setting $Z^1 = Z$, and $Z^2 = 1 - Z$ for $Z\in[0, 1]$.  This leads to $p(x) = \int_0^1 p(x\g z)p(z)\dz$, where now $\dz$ is Lebesgue measure.

\subsubsection{Definitions}
Let $U\sim g$ be a any continuous random vector $(U^1,\ldots,U^M)$ in $\RM$ having \iid components $U^i\sim g_c$, each with pdf symmetric about zero.  Then for $\pi\in\RM$, $\sigma>0$ set
\begin{align*}
  Z &= F(U; \pi, \sigma) := \softmax(\sigma \pi + \sigma U),
\end{align*}
where for $t \in \RM$, the $m^{th}$ component
\begin{align*}
  \softmax(t)^m :&= \left\{ 
  \begin{matrix}
    &\exp\left\{t^m\right\} \big/ \left(1 + \sum_{m=1}^M \exp\left\{ t^m \right\}\right),\quad 0 < m \leq M - 1\\
    &1 \big/ \left(1 + \sum_{m=1}^M \exp\left\{ t^m \right\}\right),\quad m = M.
  \end{matrix}
  \right.
\end{align*}
Note that we use the ``centered'' version of the softmax, which is a diffeomorphism from $\RM\to\Delta^M$.  A cubature constant-probability-point quadrature scheme for this $p(z)$ is constructed in section~\ref{subsection:cubature-convergence}. 

Our $M=2$ identification $\Delta^M \leftrightarrow [0, 1]$ allows replacing the softmax with a sigmoid (listing~\ref{listing:sigmoid-normal}).
\begin{align*}
  Z &= F(U;\pi, \sigma) := \sigmoid(\sigma\pi + \sigma U),\qquad
  \sigmoid(t) := \frac{\exp\left\{ t \right\}}{1 + \exp\left\{ t \right\}}.
\end{align*}
If the quantiles $\tilde\nu_n$ of $g(u)$ are available, their images $\nu_n := \sigmoid(\sigma\pi + \sigma\tilde\nu_n)$ are quantiles of $p(z)$ suitable for the scheme~\ref{definition:midpoint-quad-bounded}.

\subsubsection{Interpretation of the Parameterization}
The parameters $\pi$ and $\sigma$ controlling the mixture weights $Z^m$ have simple interpretations.

First, $\pi^i>\pi^j$ implies $Z^i$ is likely to be larger than $Z^j$, \emph{independent} of $\sigma$.  To show this, let $\Phi_s(t)$ be the probability that the sum of $s$ \iid component draws $U^k\sim g_c$ are greater than $t$.  Then, using the symmetry of $U^k$, if $i, j \neq M$,
\begin{align*}
  \rmP[Z^i > Z^j] &= \rmP[\pi^i + U^i > \pi^j + U^j] = \rmP[U^i + U^j > \pi^j - \pi^i] = \Phi_2\left(\pi^i - \pi^j\right),
\end{align*}
and if $i\neq M$,
\begin{align*}
  \rmP[Z^i > Z^M] &= \rmP[\pi^i + U^i > 0] = \rmP[U^i > -\pi^i] = \Phi_1(\pi^i).
\end{align*}

Second, the probability that one component $Z^i$ gets the majority of the weight is monotonically increasing in $\sigma$.  To see this, let $\xi^i := \pi^i + U^i$, $i=0,\ldots,M-1$, and $\xi^M = 0$.  Furthermore, let $\ell\in\left\{ 0,\ldots,M \right\}$ be the index of the maximal $\xi^i$; $\xi^\ell > \xi^i$, $i\neq\ell$.  Note that one component of $Z$ has more than 1/2 the weight as soon as
\begin{align*}
  1 > \sum_{i\neq\ell}e^{\sigma(\xi^i - \xi^\ell)}.
\end{align*}
Since all the terms $\xi^i - \xi^\ell$ are negative, the right hand side is monotonically decreasing in $\sigma$, which proves our claim.

\subsection{Reparameterizable quadrature compounds}

Here we define a quadrature compound, reparameterizable with respect to a parameter $\lambda$.
\begin{definition}[Reparameterizable Quadrature Compounds]
  \label{definition:reparameterizable-qc}
  For continuous density $p$ supported in $\calZ$, let $(\calW_N, p)$ be a quadrature scheme with weights $w_n$ independent of $\lambda$.
  Let $r$ be a probability density independent of $\lambda$ and supported on $\calV$, and for every $z\in\calZ$ let $H_z$ be a diffeomorphism $:\calV\to\calX$.
  We define the \emph{reparameterizable quadrature compound} as the density
  \begin{align*}
    q_N(x) :&= \sum_{n=1}^N p(x\g z_n) w_n,
  \end{align*}
  where $p(x\g z) := [\pushfwd{(H_z)}r](x)$.
\end{definition}

As a consequence of definition~\ref{definition:reparameterizable-qc}, samples from a reparameterizable QC are generated as
\begin{enumerate}
  \item draw $Z\sim\catw$
  \item draw $V\sim r(v)$
  \item set $X = H_{Z}(V)$.
\end{enumerate}
Our samples $X$ are thus written as a diffeomorphic transformation of $Z$ and $V$, and (given sufficient regularity) are therefore reparameterized with respect to $\lambda$ as in theorem~\ref{theorem:reparameterizable-conditions}.  Thus, the finite sample estimate
\begin{align*}
  S_K :&= \frac{1}{K}\sum_{k=1}^K \varphi(X_k), \quad X_k = H_{Z_k}(V_k),\quad Z_k\sim\catw,\quad V_k\sim r(v)
\end{align*}
is a candidate for the loss in a gradient based scheme using auto-differention.

Moreover, since $(\calW, p)$ is a quadrature scheme,
\begin{align*}
  q_N(x)\to \int p(x\g z)p(z)\dz,\quad\mbox{as}\quad N\to\infty.
\end{align*}

\subsection{Diffeomixtures and the \texttt{VectorDiffeomixture}}

\begin{definition}[Diffeomixture]
  \label{def:diffeomixture}
  A diffeomixture is a reparameterizable QC (definition~\ref{definition:reparameterizable-qc}) where $Z\in\Delta^M$.
\end{definition}

Definition~\ref{def:diffeomixture} is quite general.  Below we make an additional specialization leading to the \emph{vector diffeomixture} (VDM): Let $H_z(V)$ be a convex combination of positive-definite location-scale transformations.  That is, given the set of parameters $\left\{ (\mu_m, L_m) \right\}_{m=0}^M$, where $\mu_m\in\Rd$, and $L_m\in\Rdxd$ are positive definite, set
\begin{align*}
  X = H_Z(V) = \left(\sum_m Z^m\mu^m\right) + \left( \sum_m Z^m L^m \right)V, \quad Z\in\Delta^M.
\end{align*}
This means our final random variable $X\sim q(x)$ is an affine transformation of $V$ with coefficients equal to a (random) convex combination of the components $(\mu^m, L^m)$.

We sample $X$ from a VDM with the steps
\begin{enumerate}
  \item draw $Z\in\RM\sim \catw$
  \item draw $V\in\Rd\sim r$
  \item set $X = \left( \sum_m Z^m\mu^m \right) + \left( \sum_m Z^m L^m \right)V$, where $Z := \softmax(\sigma\pi + \sigma U)$.
\end{enumerate}
The probability density will be
\begin{align*}
  q(x) &= \sum_n p(x\g z_n) w_n.
\end{align*}

As an important example, consider $Z = F(U) = \softmax(\sigma\pi + \sigma U)$ the softmax mixture weight, and use the cubature (or quantile midpoint if $M=2$) quadrature scheme~\ref{definition:cubature-constant-probability-point-scheme},~\ref{definition:midpoint-quad-bounded}.
This means, as $N\to\infty$,
\begin{align}
  \label{align:vdm-approx}
  q(x) &\to \int_{\RM} p(x\g\softmax(\sigma\pi + \sigma u))g(u)\du.
\end{align}

Since derivatives of $u\mapsto \softmax(\sigma\pi + \sigma u)$ increase in magnitude with larger $\sigma$, the results of section~\ref{section:approximation} indicate that if $\sigma$ is large, so too should be $N$, or else the approximation \eqref{align:vdm-approx} will no longer hold.

\begin{figure}
  \centering
  \includegraphics[width=0.46\textwidth]{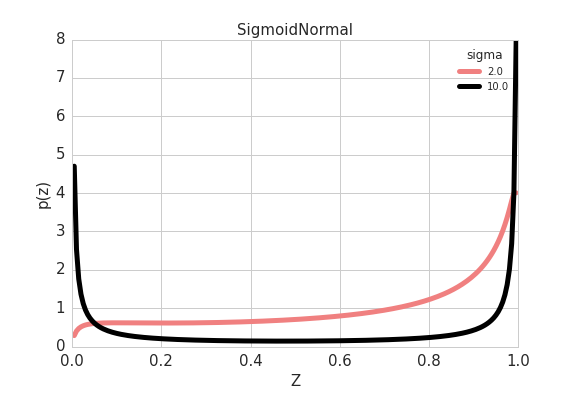}
  \includegraphics[width=0.46\textwidth]{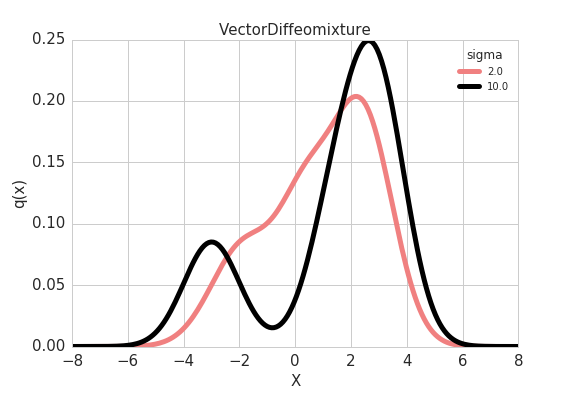}
  \caption{Left:  SigmoidNormal with $\pi=0.5$, $\sigma\in\left\{ 2, 10 \right\}$.  Right:  VectorDiffeomixture of two Normals (at $\mu=\pm3$) built with the SigmoidNormal.}
  \label{fig:vdm}
\end{figure}

Figure~\ref{fig:vdm} shows the density function of two-component ($M=2$) VDM's with different values of $\sigma$, along with a sigmoid-normal with the same $\sigma$.

The VDM has been implemented in TensorFlow.  Code for the VDM used to generate figure~\ref{fig:vdm} is shown here in listing~\ref{listing:vdm}
\begin{lstlisting}[language=Python, caption=VectorDiffeomixture, label={listing:vdm}, float=h]
  import tensorflow as tf
  import numpy as np
  ds = tf.contrib.distributions
 
  sigma = np.array([2.0, 10.0], dtype=np.float32).reshape(-1, 1)
  pi = np.array([0.5, 0.5], dtype=np.float32).reshape(-1, 1)

  vdm = ds.VectorDiffeomixture(
      mix_loc=sigma * pi, mix_scale=sigma,
      distribution=ds.Normal(0., 1.),
      loc=[[3.], [-3.]],
      scale=[
          tf.linalg.LinearOperatorDiag([1.]),
          tf.linalg.LinearOperatorDiag([1.]),
      ]
  )
\end{lstlisting}

\section{Comparison of Different Quadrature Schemes}
\label{section:approximation}
Here we analyze the schemes from section~\ref{section:qc-trick} as well as two new ones.  Convergence results and numerical experiments are presented.

A scheme with $N$ points requires evaluation of $p(x\g z_n)$ at $N$ different points $z_1,\ldots,z_N$.  In order to keep computational complexity down, we prefer schemes that work well when $N$ is not too large (around 10 - 20 worked well in experiments).  

\begin{figure}
  \centering
  \includegraphics[width=0.46\textwidth]{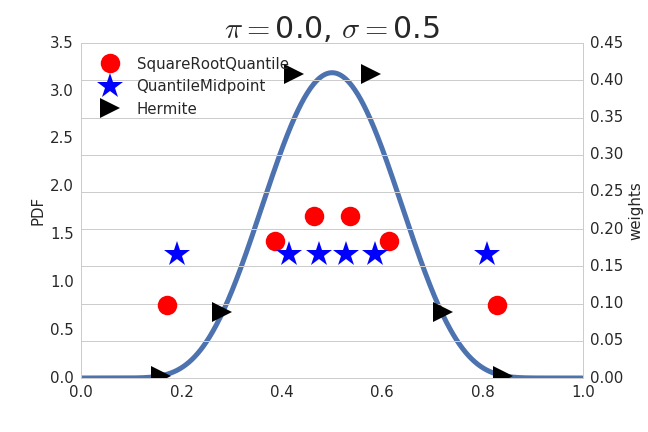}
  \includegraphics[width=0.46\textwidth]{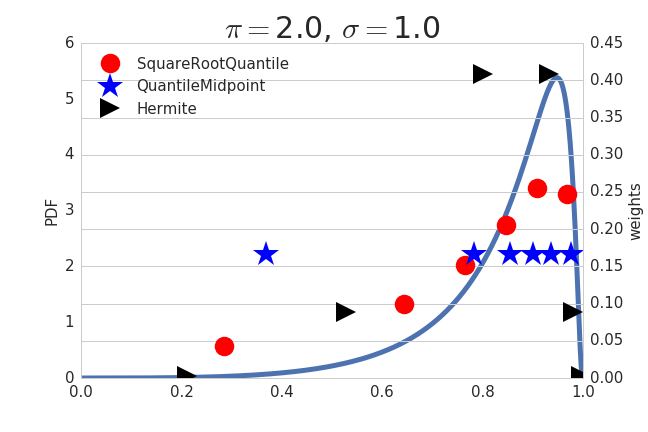} \\
  \includegraphics[width=0.46\textwidth]{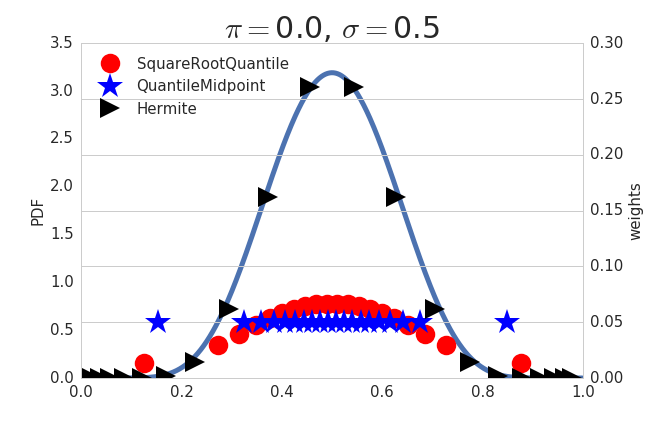}
  \includegraphics[width=0.46\textwidth]{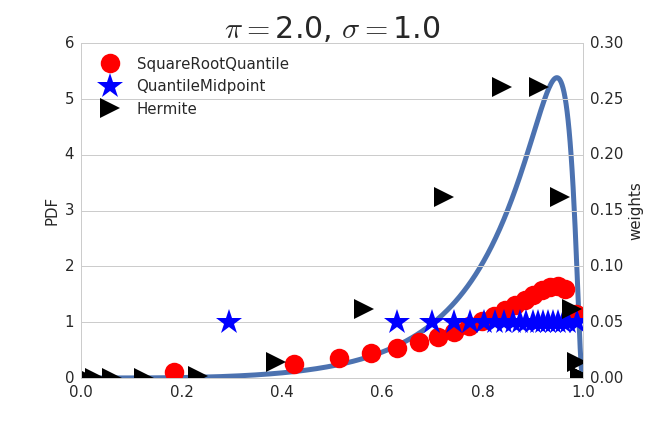}
  \caption{Quadrature points/weights for three schemes for the Sigmoid Normal distribution for the random variable $Z = F(U) = \sigmoid(\sigma\pi + \sigma U)$, $U\sim\calN(0, 1)$, along with the PDF of the sigmoid normal.  Top row was generated with 6 quadrature points, bottom with 20.  Comparing the square root quantile (section~\ref{subsection:square-root-quantile}) and quantile midpoint schemes we see that the square root quantile has fewer large gaps between grid points.  The Hermite scheme is similar when $\pi=0$, although it does put most of its points away from the mode of $p(z)$.  When $\pi=2$ the Hermite scheme has points/weights that seem incorrectly positioned w.r.t.\ the mode of $p(z)$.  This is not a bug, but instead is due to the fact that the mode of $Z$ is \emph{not} the translation of the mode of $U$ by $F$.}
  \label{fig:quad-points}
\end{figure}

\subsection{The quantile midpoint scheme for bounded intervals}
\label{subsection:quantile-convergence}
The quantile midpoint scheme puts more weight in high density regions of $p(z)$ (figure~\ref{fig:quad-points}), leading to low overall error (table~\ref{table:main-results}).  Furthermore it leads to a reparameterizable QC.  We thus consider this the best overall scheme in this paper.

This proposition shows that the scheme~\ref{definition:midpoint-quad-bounded} enjoys the same convergence properties as traditional midpoint integration.  This is not surprising. 
\begin{proposition}
  \label{proposition:quantile-midpoint-convergence}
  Suppose for every $x$, $z\mapsto p(x\g z)$ along with its first two derivatives are bounded.  Suppose $p(z)$ along with its first derivative are bounded above, and $p(z)$ is bounded below.  Then, with $q_N(x)$ a quadrature compound using the quantile midpoint scheme, there exist $C, C'$ independent of $N$ such that
  \begin{align*}
    |p(x) - q_N(x)| \leq C\sum_{n=1}^N (\nu_{N, n} - \nu_{N, n - 1})^3\leq \frac{C'}{N^2}.
  \end{align*}
\end{proposition}
\begin{proof}
  With $\delta_{N,n} = (\nu_{N,n-1}, \nu_{N,n})$, we can write
\begin{align*}
  p(x) - q_N(x)
  &= \sum_{n=1}^N\int_{\delta_{N,n}} \left[ p(x\g z) - p(x\g z_n) \right]p(z)\dz.
\end{align*}
We will first show that each summand is $O( (\nu_{N, n}- \nu_{N, n-1})^3)$, which gives the first inequality.

Let $\gamma(z):= p(x\g z)$, then, for $z\in\delta_n$ we have the Taylor expansions
\begin{align*}
  \gamma(z) &= \gamma(z_n) + \gamma'(z_n) (z - z_n) + \gamma''(\xi(z))(z - z_n)^2/2,\quad \xi(z)\in\delta_n,\\
  p(z) &= p(z_n) + p'(\eta(z))(z - z_n),\quad \eta(z)\in\delta_n.
\end{align*}
This gives us
\begin{align*}
  \int_{\delta_n} &\left[ \gamma(z) - \gamma(z_n) \right]p(z)\dz \\
  &= \gamma'(z_n)\int_{\delta_n}(z - z_n)p(z)\dz + \int_{\delta_n}\gamma''(\xi(z))\frac{(z - z_n)^2}{2} p(z)\dz \\
  &= \gamma'(z_n)p(z_n)\int_{\delta_n}(z - z_n)\dz + \gamma'(z_n)\int_{\delta_n}(z-z_n)^2 p'(\eta(z))\dz \\
  &\qquad + \int_{\delta_n}\gamma''(\xi(z))\frac{(z - z_n)^2}{2} p(z)\dz.
\end{align*}
The first term above vanishes because $\int_{\delta_n}(z - z_n)\dz=0$.  The second two terms are both bounded by a constant times $\int_{\delta_{N,n}}(z - z_n)^2\dz \leq (\nu_{N,n}-\nu_{N,n-1})^3$, and thus we have the first inequality.

The second inequality will follow once we show $\nu_{N,n} - \nu_{N,n-1}$ is bounded by a constant times $1/N$.  This fact follows from the relation
\begin{align*}
  \frac{1}{N} &= \int_{\delta_{N,n}}p(z)\dz \geq (\nu_{N,n}-\nu_{N,n-1}) \cdot \min_z p(z).
\end{align*}
\end{proof}

\subsection{The square root quantile scheme}
\label{subsection:square-root-quantile}
By construction, placing quadrature grid points $z_n$ at quantile midpoints of $p(z)$ results in fewer points in low density regions (see e.g. figure~\ref{fig:quad-points}).  This reduces error in the important regions where $p(z)$ is large, at the possible expense of missing high value regions of $p(x\g z)$.  This can roughly be rephrased as a trade off in errors proportional to $p'(z)$ and $\p_x p(x\g z)$.  Here we argue that in some cases the correct balance is to replace the quantiles of $p$ with those from a distribution proportional to $\sqrt{p}$.  This decreases the number of grid points in regions where $p(z)$ is larger than average, and increases the number of grid points in regions where $p(z)$ is smaller than average. The major drawback of this scheme is that the weights now depend on parameters defining $p(z)$, hence it does not lead to a reparameterizable QC.  Furthermore, the quantiles of $\sqrt{p}$ are likely not readily available, and in our case a separate numerical integration had to be performed.

Consider a generic midpoint scheme with end points $a_0<a_1\cdots<a_N$ and midpoints $z_n = (a_{n-1} + a_n)/2$.  This leads to the approximation
\begin{align*}
  q_N(x) &= \sum_{n=1}^N p(x\g z_n) w_n,\qquad w_n := \int_{a_{n-1}}^{a_n}p(z)\dz.
\end{align*}
If $z\mapsto p(x\g z)$ is Lipshitz with constant $L$, and $p(z)$, $p'(z)$ are bounded above, and $p(z)$ is bounded below, we have (similar to the proof of proposition~\ref{proposition:quantile-midpoint-convergence}
\begin{align*}
  |p(x) - q_N(x)|  &\leq Lp(z_n)\sum_n \int_{a_{n-1}}^{a_n} |z - z_n|\dz + O(1/N^2).
\end{align*}
Ignoring the $O(1/N^2)$ term, we would like to choose $a_n$ to minimize $\sum_n p(z_n)\int_{a_{n-1}}^{a_n}|z - z_n|\dz = \sum_n p(z_n) (a_n - a_{n-1})^2/2$.  Instead of solving this optimization problem, we deal with the simpler task of ensuring $p(z_n)(a_n - a_{n-1})^2\leq \eps$, for some prescribed error $\eps$.  This is achieved if $a_n - a_{n-1} = \eps / \sqrt{p(z_n)}$.  On the other hand, if we let $a_n$ be the quantiles of a distribution proportional to $\sqrt{p}$, then with $C := \int \sqrt{p(z)}\dz$,
\begin{align*}
  \frac{1}{N}
  &= \frac{1}{C}\int_{a_{n-1}}^{a_n}\sqrt{p(z)}\dz
  = \frac{1}{C}\sqrt{p(z_n)} (a_n - a_{n-1}) + O(1/N^2).
\end{align*}
Ignoring the $O(1 / N^2)$ term, this means $a_n - a_{n-1} \leq \eps / \sqrt{p(z_n)} $, as soon as $N \geq C/\eps$.

\subsection{Cubature constant-probability schemes}
\label{subsection:cubature-convergence}
To start with a useful example, we will first construct a cubature scheme for the softmax mixture weights of section~\ref{subsection:sigmoid-softmax-mixture-weights}.  We then show that the cubature scheme definition~\ref{definition:cubature-constant-probability-point-scheme} in fact leads to a bone-fide quadrature scheme (e.g.\ $q_N(x)\to p(x)$ as $N\to\infty$).
Since the cubature scheme generalizes the others presented in section~\ref{section:qc-trick}, this shows that these other schemes are also quadrature schemes under the cubature scheme's relaxed assumptions on $p(x\g z)$.

Recall the softmax mixture weight $Z = \softmax(\sigma\pi + \sigma U)$, where $U = (U^1,\ldots,U^M)\subset\RM$, with each $U^m\sim g_c(u)$.  To construct a cubature scheme for $Z$, we first slice up the interval $(-\infty,\infty)$ into quantiles of $g_c(u)$:  For integers $K\geq1$ and $k\in\left\{ 0,\ldots,K \right\}$, let $\nu_k$ be the $k/K$ quantile of $g_c$.  That is, $\rmP[U\leq\nu_k] = k/K$.  Note we may have $\nu_0=-\infty$, $\nu_K=\infty$.  Now for $(k_1,\ldots,k_M)\in \left\{ 1,\ldots,K \right\}^M$ set
\begin{align*}
  \tilde\delta_{k_1,\ldots,k_M} :&= (\nu_{k_1 - 1}, \nu_{k_1})\times\cdots\times (\nu_{k_M - 1}, \nu_{k_M}).
\end{align*}
This is a partition of $\RM$ into $K^M$ regions such that $\rmP[(U^1,\ldots,U^M)\in \tilde\delta_{k_1,\ldots,k_M}] \equiv 1 / K^M$.  Re-index these into $\tilde\delta_{1,N},\ldots,\tilde\delta_{N,N}$, with $N = K^M$ and set $\delta_{N,n} := F(\tilde\delta_{N,n})$, with $F(u) := \softmax(\sigma\pi + \sigma u)$.

\begin{proposition}
  \label{proposition:softmax-cubature}
$\delta_{n,N}$ constructed as above is a constant probability cubature scheme (definition~\ref{definition:cubature-constant-probability-point-scheme}).
\end{proposition}
\begin{proof}
Since by construction,
\begin{align*}
  \rmP[Z\in\delta_{n,N}] &= \rmP[(U^1,\ldots,U^M)\in\tilde\delta_{n,N}] = 1 / N,
\end{align*}
we only need to show that for every $D > 0$, $B_{D, N} := \union\left\{ \delta_{n,N}\st \diam(\delta_{n,N}) > D \right\}$ tends to zero in $p(z)\dz$ measure.  Given $\eps > 0$, define $\calK_\eps$ to be the closed ball centered at the origin such that $\rmP[U\in \calK_\eps] = 1 - \eps$.  We will show that given $D > 0$, there exists $N'$ such that for $N > N'$, $B_{D, N}\subset \calZ\setminus F(\calK_\eps)$, then since $\rmP[Z\in \calZ\setminus F(\calK_\eps)] = \eps$ the proof will be complete.  To that end, choose $N$ such that $N^{1/M} > (M\sigma) / (D \min_{u\in\calK_\eps}g(u))$.  Then for $u, \tilde u\in \tilde\delta_{n,N}\subset \calK_\eps$, and $m\in\left\{ 1,\ldots,M \right\}$, the $m^{th}$ components $u^m$, $\tilde u^m$ lie between two quantiles of $g_c$, $(\nu_{k_\ell}, \nu_{k_{\ell+1}})$ having $g_c$-measure $1 / N^{1/M}$. This means
\begin{align*}
  \frac{1}{N^{1/M}} &= \int_{\nu_{k_\ell}}^{\nu_{k_{\ell+1}}} g_c(u)\du \geq \min_{u\in\calK_\eps}g_c(u)\cdot |u^m - \tilde u^m|,
\end{align*}
from which it follows $|u^m - \tilde u^m| < D / (M\sigma)$.  Since the same holds for every $m = 1,\ldots,M$, we have $\|u - \tilde u\| \leq D / \sigma$.
Now consider the points $z = F(u)$, $\tilde z = F(\tilde u)$.  Note that $\|\nabla F\|_\infty < \sigma$, and thus, with $f(t) := F(u + t (\tilde u - u) / \|\tilde u - u\|)$,
\begin{align*}
  |\tilde z - z|
  = \left|f(\|\tilde u - u\|) - f(0)\right|
  &= \left| \int_0^{\|\tilde u - u\|} f'(t)\dt \right|
  \leq \sigma \|\tilde u - u\|
  \leq D,
\end{align*}
which shows $\tilde z, z\notin B_{D, N}$.  Since $\tilde z, z\in F(\calK_\eps)$ were arbitrary, we conclude $B_{D,N}\subset\calZ\setminus F(\calK_\eps)$ and the proof is complete.
\end{proof}

\begin{proposition}
  \label{proposition:cubature-convergence}
  Assume $z\mapsto p(x\g z)$ is uniformly continuous and bounded.  Then the cubature constant-probability points scheme~\ref{definition:cubature-constant-probability-point-scheme} is a convergent quadrature scheme (in the sense of~\ref{definition:quadrature-scheme}) that generates a reparameterizable QC distribution.
\end{proposition}
\begin{proof}
  The cubature scheme has constant weights $w_n\equiv 1/N$, so for any fixed $N$ the distribution $N^{-1}\sum_n p(x\g z_n)$ will be reparameterizable.  To show convergence, let $\gamma:\calZ\to\Rone$ be uniformly continuous and bounded, and  write
  \begin{align*}
    \int_\calZ &\gamma(z)p(z)\dz - \frac{1}{N}\sum_{n=1}^N \gamma(z_{N,n})
    = \int_\calZ \sum_{n=1}^N \one_{\delta_{N,n}(z)}\left[ \gamma(z) - \gamma(z_{N,n}) \right]p(z)\dz.
  \end{align*}
  Equality holds because $\rmP[Z\in\delta_{N,n}]\equiv 1 / N$.  Denote the integrand by $f_N(z)$.  Given $\eps > 0$, we will find $N'$ large enough such that $\left|\int_S f_N(z)p(z)\dz\right| < \eps$ for all $N > N'$, which will prove the proposition.
  
  First note that the uniform continuity of $\gamma$ implies there exists $\eps' > 0$ such $|\gamma(z) - \gamma(z_{N,n})| < \eps / 2$ for every $n$ such that $\diam(\delta_{n,N}) < \eps'$.  Second, the boundedness of $\gamma$ and the constant-probability-point scheme hypothesis regarding $B_{D,N}$ implies that there exists $N' > 0$ such that for $N > N'$, the set of $\delta_{n,N}$ with diameter greater than $\eps'$ (this set is called $B_{\eps', N}$) has combined measure less than $\eps / (4\|\gamma\|_\infty)$.  So, for $N > N'$,
  \begin{align*}
    \left| \int_\calZ f_N(z)p(z)\dz \right|
    &\leq \int_{B_{\eps', N}} \left| f_N(z)  \right| p(z)\dz +  \int_{\calZ\setminus B_{\eps', N}} \left| f_N(z)\right| p(z)\dz \\
    &\leq 2\|\gamma\|_\infty \frac{\eps}{4\|\gamma\|_\infty} + \frac{\eps}{2} \rmP[Z\in\calZ\setminus B_{\eps', N}] \\
    & \leq \frac{\eps}{2} + \frac{\eps}{2}.
  \end{align*}
\end{proof}

\subsection{Pushforwards of schemes}
\label{subsection:pushforwards-of-schemes}
Consider the case where $p(z) = (\pushfwd{F}g)(z)$ for some diffeomorphism $F$ as in section~\ref{section:transformed-distributions}.  If we have a quadrature scheme $\left\{ (u_1, w_1),\ldots, (u_N, w_N) \right\}$ appropriate for $g(u)$, we can ``push this scheme forward'' to one for $p(z)$.  Write
\begin{align*}
  p(x) &= \int p(x\g z)p(z)\dz = \int p(x\g F(u)) g(u)\du \approx \sum_n p(x\g F(u_n)) w_n =: q_N(x),
\end{align*}
which gives us a quadrature scheme for $p(z)$ with points $z_n := F(u_n)$ and weights $w_n$.  This formalism allows $F$ to depend on parameters $\lambda$, and through these parameters we adjust the distribution $p(z)$.  If in addition $p(x\g z)$ is continuous in $x$ (and differentiable in $\lambda$), and the $w_n$ are independent of $\lambda$, $q_N$ will be a QC reparameterizable with respect to $\lambda$.

\subsection{Gaussian quadrature}
\label{subsection:guassian-quadrature}
\begin{figure}
  \centering
  \includegraphics[width=0.46\textwidth]{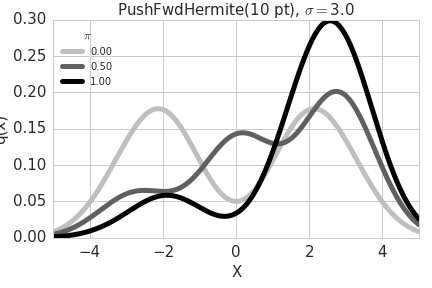}
  \includegraphics[width=0.46\textwidth]{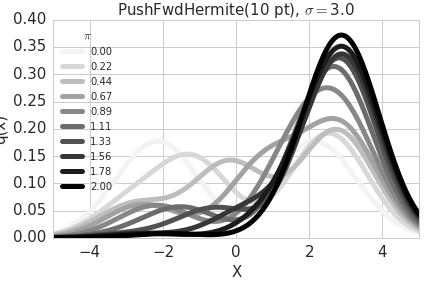} \\
  \includegraphics[width=0.46\textwidth]{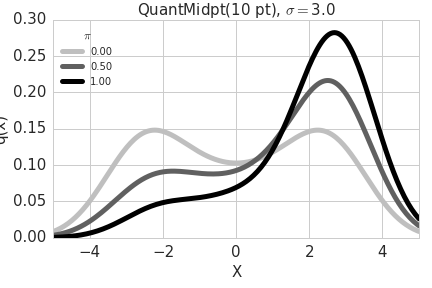}
  \includegraphics[width=0.46\textwidth]{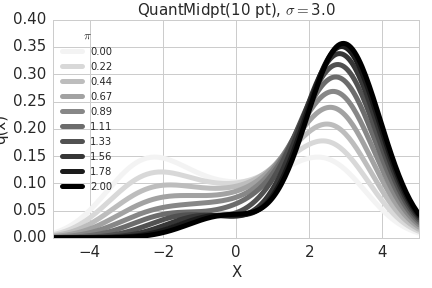}
  \caption{PDF for the 2-component VDM shown also in figure~\ref{fig:vdm-smooth}, with slightly different $\pi$, $\sigma$, and 10 quadrature points.  Top:  10 points are from the pushforward scheme based on Gauss-Hermite quadrature.  Bottom:  Using the quantile midpoint scheme.  Comparing the two, the Gauss-Hermite scheme results in spurious bumps in the center, due to sparcity of Gauss-Hermite points near the center.}
  \label{fig:hermite-bad}
\end{figure}

Gaussian quadrature is a general technique for estimating integrals of the form $\int \gamma(u)g(u)\du$, for various weight functions $g(u)$ (not just Gaussian distributions).  It is optimal in some situations, and hence it is tempting (and sometimes appropriate) to use in the pushfoward scheme of section~\ref{subsection:pushforwards-of-schemes}.  We briefly review Gaussian quadrature, and discuss its pros and cons for QC distributions.  A surprising result is that Gauss-Hermite quadrature gives terrible results until the number of quadrature points is quite high (around 50 in our experiments).  Before going into details, see figure~\ref{fig:hermite-bad}.

First let us consider a case where Gaussian quadrature is appropriate.  Suppose $g(u) = g(u;\alpha,\beta) \propto x^\alpha (1-x)^\beta$ is the beta distribution.  We may then find a sequence of polynomials $v^{\alpha,\beta}_n$, such that
\begin{align*}
  \int_0^1 g(u;\alpha,\beta)v^{\alpha,\beta}_n(u) v^{\alpha,\beta}_m(u)\du
  &= \left\{  
  \begin{matrix}
    1,&\quad m=n,\\
    0,&\quad m\neq n.
  \end{matrix}
  \right.
\end{align*}
Given $N\in\Nat$, designate the roots of $v^{\alpha,\beta}_N$, as $u_{N,n}$.  We then have (see~\cite{stoer-numerical-analysis})
\begin{align}
  \label{align:jacobi-quad-result}
  \begin{split}
    p(x)
    = \int_0^1 p(x\g F(u)) g(u;\alpha,\beta)\du
    &= \sum_{n=1}^N p(x\g F(u_{N,n})) g(u_{N,n};\alpha,\beta)
    + e_N(p(x\g\cdot)), \\
    |e_N(p(x\g\cdot))| &\leq C\frac{\max_{0\leq\xi\leq1}|\p^{2N}_\xi p(x\g F(\xi))|}{(2N)!},
  \end{split}
\end{align}
For some $C>0$ independent of $p(x\g\cdot)$, and $N$.  We thus have (for sufficintly smooth $p(x\g\cdot)$), quite fast convergence.  The $v^{\alpha,\beta}$ are the (re-scaled) Jacobi polynomials, and the roots/weights are available (after rescaling) through many numerical packages.  There is a key limitation though, in that the weights will depend on $(\alpha,\beta)$, and therefore this will not yield a reparameterized QC with respect to $(\alpha,\beta)$.  Also note that the convergence will be less impressive if $p(x\g F(u))$ was not sufficiently smooth.

Keeping in mind our desire to use methods based on fewer quadrature points $N$, we should not be too excited about the error bound in \eqref{align:jacobi-quad-result}.  For example, if $p = \pushfwd{F}g$, with $F(u) = \sigmoid(\sigma\pi + \sigma u)$, then $\p_u^{2N} p(x\g F(u))\sim O(\sigma^{2N})$ which can be comparable to $(2N)!$ for smaller $N$.  In the case of beta distribution $g(u; \alpha,\beta)$, an error estimate converging in $N$ and requiring only one derivative is achieved in~\cite{gauss-jacobi-error}.  Let $r\in\Nat$, then
\begin{align}
  \label{align:gauss-jacobi-error}
  |e_N(p(x\g\cdot))| &\leq \frac{C}{N^r} \int_0^1 |v^{\alpha,\beta}(u) \p_u^r p(x\g F(u)) \p^r_u \sqrt{(1 - u)u}| \du.
\end{align}

Next, we consider the case where $g(u)$ is Gaussian, and argue that Gaussian quadrature (using Hermite polynomials) is not appropriate for a QC.  This may come as a surprise.  The key difficulty is that the domain in question is the entire real line $(-\infty,\infty)$, and thus, as $N$ grows, the quadrature points $u_{N,n}$ are placed further and further from the origin.  See e.g. figure~\ref{fig:quad-points}, which shows the points $z_n = \sigmoid(\sigma\pi + \sigma u_{N,n})$ for the Hermite scheme.  Most of the points $z_n$ are near are near the boundary where there is little mass in $p(z)$, and only a few points are near the mode of $p(z)$.  In~\cite{gauss-unbounded}, the bound \eqref{align:gauss-jacobi-error} is compared with a similar one when $g(u)$ is Gaussian.  In this case,
\begin{align}
  \label{align:gauss-hermite-error}
  |e_N(p(x\g\cdot))| &\leq \frac{C}{N^{1/6}} \int |\p_u p(x\g F(u))| g(u)\du.
\end{align}
Moreover, this error is shown to be tight with a quite tame (but only piecewise smooth) function.  Thus we have gone from a hopeful $1 / (2N)!$ to a disimal $1 / N^{1/6}$.

\medskip

\subsection{Numerical comparison of four schemes}

We constructed a 2-component vector diffeomixture $q_N(x)$ (see section~\ref{section:reparameterizable-qc-and-diffeomixture}), with each component a 10 dimensional multivariate Normal, centered at $\pm (\mu,\ldots,\mu)\in\Rone^{10}$.  We used various quadrature schemes, and compared to a reference VDM $p(x)$ built using a scheme with 150 points (with 150 points, all schemes gave the same result).  Comparison was done using KL divergence and total variation for various parameters.  We used every combination of mixture component bias $\pi\in\left\{ 0.0, 0.5, 1.0, 1.5, 2.5 \right\}$, mixture scale $\sigma\in\left\{ 2, 5 \right\}$, number of quadrature points $N\in \left\{ 5, 10, 20, 50 \right\}$, and component center magnitude $\mu\in\left\{ 2, 4 \right\}$.
The mean error over all sweep parameters is shown in table~\ref{table:main-results}.
\begin{table}[h!]
  \label{table:main-results}
  \centering
\begin{tabular}{|l||c|c|c|}
 \hline 
            & SqrtQuantMidpt & QuantMidpt & PushFwdHermite \\
 \hline\hline 
 $\KL{q_N}{p}$ & 0.04 & 0.07 & 0.31 \\
 \hline 
 $\KL{p}{q_N}$ & 0.06 & 0.13 & 1.13 \\
 \hline 
 $\TV{p}{q_N}$ & 0.06 & 0.07 & 0.21 \\
 \hline 
\end{tabular}
\caption{\texttt{SqrtQuantMidpt} is the scheme based on quantiles of the distribution proportional to $\sqrt{p}$ that we present in section~\ref{subsection:square-root-quantile}.  This generally had the lowest error.  The quantile midpoint scheme \texttt{QuantMidpt} from defintion~\ref{definition:midpoint-quad-bounded} often performed almost as well.  The pushforward of Gauss-Hermite quadrature \texttt{PushFwdHermite} (see section~\ref{subsection:guassian-quadrature}) ususally did worse, and often much worse.}
\end{table}

Considering only total variation (the KL divergences give similar results), we compare average error for different number of quadrature points $N$.k
\begin{table}[h!]
  \label{table:different-quad-points-results}
  \centering
\begin{tabular}{|l||c|c|c|}
 \hline 
            & SqrtQuantMidpt & QuantMidpt & PushFwdHermite \\
 \hline\hline 
 $N = 5$ & 0.19 & 0.19 & 0.34 \\
 \hline 
 $N = 10$ & 0.04 & 0.06 & 0.23 \\
 \hline 
 $N = 20$ & 0.01 & 0.02 & 0.18 \\
 \hline 
 $N = 50$ & 0.00 & 0.00 & 0.10 \\
 \hline 
\end{tabular}
\caption{Same comparision as table~\ref{table:main-results}, but limited to total variation distance, and grouped by number of quadrature points $N$.  This shows much faster convergence for the quantile schemes than the Hermite.}
\end{table}



\begin{thebibliography}{10}

\bibitem{Blei06correlatedtopic}
D.~M. Blei and J.~D. Lafferty.
\newblock Correlated topic models.
\newblock In {\em In Proceedings of the 23rd International Conference on
  Machine Learning}, pages 113--120. MIT Press, 2006.

\bibitem{tensorflow}
M.~A. et. al.
\newblock {TensorFlow}: Large-scale machine learning on heterogeneous systems,
  2015.
\newblock Software available from tensorflow.org.

\bibitem{fernandes-gauss-arbitrary-positive}
A.~D. Fernandes and W.~R. Atchley.
\newblock Gaussian quadrature formulae for arbitrary positive measures.
\newblock {\em Evolutionary Bioinformatics}, 2:251--259, 2006.

\bibitem{folland}
G.~B. Folland.
\newblock {\em Real analysis. Modern techniques and their applications. 2nd
  ed.}
\newblock Wiley, 2nd ed. edition, 2007.

\bibitem{fu-gradient-estimation}
M.~Fu.
\newblock {\em Simulation}, volume~13 of {\em Handbook in Operations Research
  and Management Science}.
\newblock North Holland, 2006.

\bibitem{gumbel-softmax}
E.~{Jang}, S.~{Gu}, and B.~{Poole}.
\newblock {Categorical Reparameterization with Gumbel-Softmax}.
\newblock {\em ArXiv e-prints}, Nov. 2016.

\bibitem{scipy-special}
E.~Jones, T.~Oliphant, P.~Peterson, et~al.
\newblock {SciPy}: Open source scientific tools for {Python}. {S}pecial
  functions package., 2001--.

\bibitem{jordan-conjugate}
M.~Jordan.
\newblock The exponential family: Conjugate priors, 2010.

\bibitem{kingma-welling}
D.~P. {Kingma} and M.~{Welling}.
\newblock {Auto-Encoding Variational Bayes}.
\newblock {\em ArXiv e-prints}, Dec. 2013.

\bibitem{concrete}
C.~J. {Maddison}, A.~{Mnih}, and Y.~{Whye Teh}.
\newblock {The Concrete Distribution: A Continuous Relaxation of Discrete
  Random Variables}.
\newblock {\em ArXiv e-prints}, Nov. 2016.

\bibitem{gauss-jacobi-error}
G.~Mastroianni.
\newblock {Generalized Christoffel functions and error of positive quadrature}.
\newblock {\em Numer. Algorithms}, 10, 1995.

\bibitem{naesseth-reparameterizable-rejection}
C.~A. {Naesseth}, F.~J.~R. {Ruiz}, S.~W. {Linderman}, and D.~M. {Blei}.
\newblock {Reparameterization Gradients through Acceptance-Rejection Sampling
  Algorithms}.
\newblock {\em ArXiv e-prints}, Oct. 2016.

\bibitem{bbvi}
R.~Ranganath, S.~Gerrish, and D.~M. Blei.
\newblock Black box variational inference.
\newblock In {\em Proceedings of the Seventeenth International Conference on
  Artificial Intelligence and Statistics, {AISTATS} 2014, Reykjavik, Iceland,
  April 22-25, 2014}, pages 814--822, 2014.

\bibitem{Robbins&Monro:1951}
H.~Robbins and S.~Monro.
\newblock A stochastic approximation method.
\newblock {\em Annals of Mathematical Statistics}, 22:400--407, 1951.

\bibitem{Schulman-2015-stochastic-computation-graphs}
J.~Schulman, N.~Heess, T.~Weber, and P.~Abbeel.
\newblock Gradient estimation using stochastic computation graphs.
\newblock In {\em Proceedings of the 28th International Conference on Neural
  Information Processing Systems - Volume 2}, NIPS'15, pages 3528--3536,
  Cambridge, MA, USA, 2015. MIT Press.

\bibitem{stoer-numerical-analysis}
J.~Stoer and R.~Bulirsch.
\newblock {\em Introduction to numerical analysis}.
\newblock Texts in applied mathematics. Springer, 2002.

\bibitem{gauss-unbounded}
B.~D. Vecchia and G.~Mastroianni.
\newblock Gaussian rules on unbounded intervals.
\newblock {\em Journal of Complexity}, 19(3):247 -- 258, 2003.
\newblock Oberwolfach Special Issue.

\end{thebibliography}
\end{document}